\documentclass[journal]{IEEEtran}
\usepackage{blindtext}
\usepackage{graphicx}

\usepackage{cite}

\ifCLASSINFOpdf
  
\else

\fi

\usepackage{multirow}

\usepackage[cmex10]{amsmath}

\usepackage{array}

\usepackage[tight,footnotesize]{subfigure}

\usepackage[caption=false,font=footnotesize]{subfig}
\usepackage{graphics}

\usepackage{textcase}
\usepackage[tablename=TABLE]{caption}
\DeclareCaptionTextFormat{up}{\MakeTextUppercase{#1}}

\begin{document}
\title{Low Temperature Homoepitaxy Of (010) $\beta$-Ga$_2$O$_3$ By Metalorganic Vapor Phase Epitaxy : Expanding The Growth Window}
 
\author{Arkka~Bhattacharyya,Praneeth Ranga, Saurav~Roy, Jonathan Ogle, Luisa Whittaker-Brooks, 
and ~Sriram~Krishnamoorthy\vspace{-0.42cm}
\thanks{Arkka Bhattacharyya, Praneeth Ranga, Saurav Roy, and Sriram Krishnamoorthy are with the Department of Electrical and Computer Engineering, The University of Utah, Salt Lake City, UT, 84112, United States of America (e-mail: a.bhattacharyya@utah.edu, sriram.krishnamoorthy@utah.edu ).}

\thanks{Jonathan Ogle and Luisa Whittaker-Brooks are with the Department of Chemistry, University of Utah, Salt Lake City, Utah 84112, USA.}

}

\markboth{}%
{Shell \MakeLowercase{\textit{et al.}}: Bare Demo of IEEEtran.cls for IEEE Journals}

\maketitle
\begin{abstract}
In this work, we report on the growth of high-mobility $\beta$-Ga$_2$O$_3$ homoepitaxial thin films grown at a temperature much lower than the conventional growth temperature window for metalorganic vapor phase epitaxy. Low-temperature $\beta$-Ga$_2$O$_3$ thin films grown at 600$^{\circ}$C on Fe-doped (010) bulk substrates exhibits remarkable crystalline quality which is evident from the measured room temperature Hall mobility of 186 cm$^2$/Vs for the unintentionally doped films. N-type doping is achieved by using Si as a dopant and a controllable doping in the range of 2$\times$10$^{16}$ - 2$\times$10$^{19}$ cm$^{-3}$ is studied. Si incorporation and activation is studied by comparing silicon concentration from secondary ion mass spectroscopy (SIMS) and electron concentration from temperature-dependent Hall measurements. The films exhibit high purity (low C and H concentrations) with very low concentration of compensating acceptors (2$\times$10$^{15}$ cm$^{-3}$) even at this growth temperature. Additionally, abrupt doping profile with forward decay of $\sim$ 5nm/dec (10 times improvement compared to what is observed for thin films grown at 810$^{\circ}$C) is demonstrated by growing at a lower temperature. 

\end{abstract}

\IEEEpeerreviewmaketitle
\section{Introduction}
\label{sec1}
\IEEEPARstart{B}{eta} 
-Ga$_2$O$_3$ is the emerging material of choice for power device applications owing to its material properties that translates to predicted Baliga figure of merit (BFOM) several times larger than SiC and GaN \cite{Higashiwaki2018}. To reach this intrinsic potential, a material with low background doping and high mobility is necessary. Carrier transport in $\beta$-Ga$_2$O$_3$ suffers from severe polar optical phonon scattering that limits its theoretical room temperature mobility to 200 cm$^2$/Vs\cite{Ma2016}. Metalorganic vapor phase epitaxy (MOVPE) has emerged as the leading growth technique that allows for the growth of epitaxial films with mobility values closer to this theoretical limit\cite{Feng2020}. So far, the homoepitaxy of $\beta$-Ga$_2$O$_3$ using MOVPE has been demonstrated with different reactor geometries, different precursors, dopants and on different substrate orientations. All the growth and doping studies done on these homoepitaxial $\beta$-Ga$_2$O$_3$ thin films are typically performed at high temperatures ($>$800$^{\circ}$C)\cite{Feng2019,Alema2017,Zhang2019,Anooz2020,Schewski2016,baldini2016,ranga2020,wagner2014}. The lower temperature regime $<$800$^{\circ}$C still remains unexplored and could help in furthering the understanding of the growth window for $\beta$-Ga$_2$O$_3$ epitaxy. 

The low temperature epitaxial growth (LT) of $\beta$-Ga$_2$O$_3$ is interesting to study especially for the development of power and high-frequency (RF) devices. Dopant incorporation, desorption and segregation in semiconductor growth are strongly correlated to the growth temperature and could be better controlled if films are grown at lower temperatures. The effect of these phenomena could be very important in the development of 2D channel RF devices such as delta-doped and modulation-doped FETs where extremely sharp doping profiles become critical. Another key application of MOVPE LT epitaxy could be the ohmic contact regrowth process to buried channel structures that could expand the capability of the MOVPE technique in terms of device processing\cite{chabak2019,xia2018}. Regrowth process if done at high temperatures could lead to dopant diffusion, creation and migration of defects in the active region that could result in channels with reduced mobility. Here, we explore the low temperature regime of the MOVPE growth window by tuning the growth temperature down to 600$^{\circ}$C. We show that $\beta$-Ga$_2$O$_3$ thin films with with electronic properties similar to thin films grown at higher temperatures are achievable. This allows us to expand the growth window of MOVPE by almost more than 200$^{\circ}$C. Furthermore, we demonstrate sharp silicon doping profile enabled by low temperature epitaxy.  

\begin{figure*}
\includegraphics[width=7in,height=6cm, keepaspectratio]{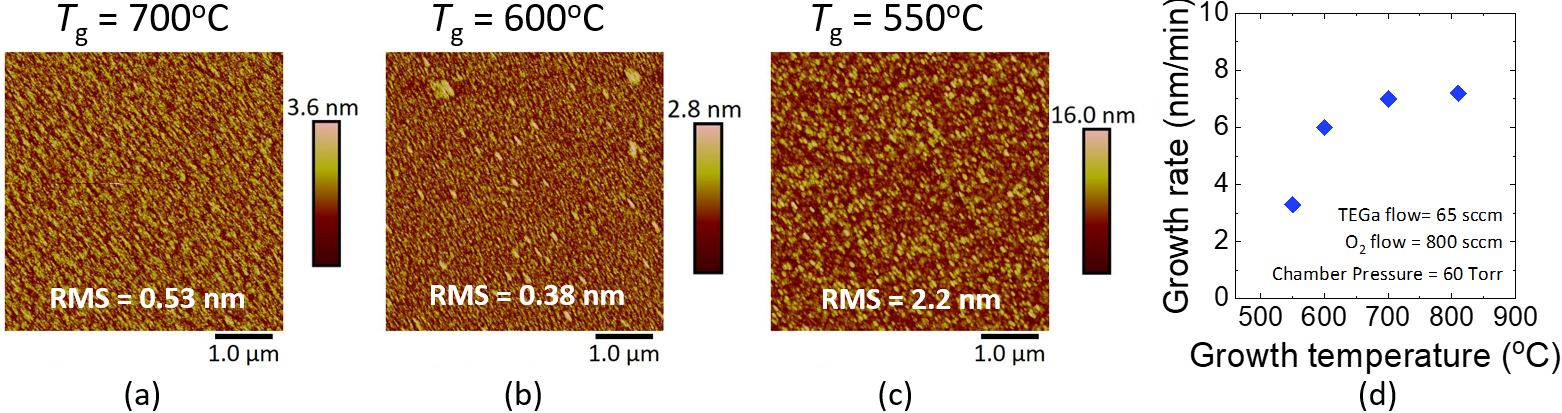}
\caption{5$\mu$m $\times$ 5$\mu$m AFM (Bruker Dimension Icon) scans of the $\beta$-Ga$_2$O$_3$ thin films grown at substrate temperatures of (a) 700$^{\circ}$C, (b) 600$^{\circ}$C and (c) 550$^{\circ}$C. (d) Growth rate as a function of the growth temperature. }
\vspace{-0.15cm}
\end{figure*}

$\beta$-Ga$_2$O$_3$ thin films were grown on Fe-doped (010) semi-insulating bulk substrates (Novel Crystal Technology, Japan) in an MOVPE reactor (Agnitron Agilis) using triethylgallium (TEGa), O$_2$ and diluted silane as precursors and argon as the carrier gas. Growths were carried out at substrate temperatures ($T_g$) of 550 - 810$^{\circ}$C.  Prior to loading the substrates into the growth chamber, the substrates were cleaned with acetone, methanol and DI water followed by a 30 min dip in a diluted HF solution\cite{Feng2020}. Films with thicknesses 400-1000 nm were grown. Uniformity of the doping profile were characterized by capacitance-voltage (CV) measurements on lateral Schottky diode structures with Ni/Au Schottky pads and Ti/Au Ohmic contacts.


The films were grown at growth temperatures of $T_g$ = 550$^{\circ}$C, 600$^{\circ}$C, 700$^{\circ}$C and 810$^{\circ}$C. The precursor flow values were set to 65 sccm, 800 sccm, 1100 sccm for TEGa, O$_2$, Ar respectively and the chamber pressure was maintained at 60 Torr. By keeping all other growth conditions identical, $T_g$ was the only growth parameter varied . Figure 1 shows the atomic force microscopy (AFM) scans of the  surface morphology of representative films grown at (a) 700$^{\circ}$C, (b) 600$^{\circ}$C and (c) 550$^{\circ}$C. The films grown at $T_g$ = 600$^{\circ}$C and 700$^{\circ}$C show extremely smooth surfaces with RMS roughness of 0.38nm and 0.53 nm respectively. The film grown at $T_g$ = 550$^{\circ}$C show a rougher surface morphology (RMS roughness = 2.2nm) indicating the onset of enhanced nucleation because of the reduced Ga adatom mobility.

The step-flow growth condition could be achieved because the diffusion coefficient of Ga adatoms are much higher on a Ga$_2$O$_3$ surface as compared to that on a GaN surface which, along with ammonia cracking constraints, limits GaN growth to higher temperatures. Early modeling and experimental work reveal that Ga adatom surface diffusivity on a Ga$_2$O$_3$ substrate could be 6 orders higher than in GaN \cite{Schewski2016}. The growth of homoepitaxial Ga$_2$O$_3$ films at lower temperatures is further supported by the fact that the cracking temperature of the precursor (TEGa) is lower than the growth temperatures used in this work\cite{miller2018epitaxial}.

Figure 1(d) shows the growth rate of the films with different $T_g$. Growth rate was measured and corroborated using SIMS and cross-sectional scanning electron microscopy (SEM) imaging of co-loaded c-plane sapphire substrates. The films grown at $T_g$ = 700$^{\circ}$C and 810$^{\circ}$C showed similar growth rate of $\sim$ 7nm/min indicating, in this regime, the growth is mass transport limited. The films grown at $T_g$ = 550$^{\circ}$C and 600$^{\circ}$C show a reduced growth rate of $\sim$ 3nm/min and $\sim$ 6nm/min respectively. This indicates that the growth rate is now surface kinetics limited. Therefore, further low temperature growth experiments were performed on films grown at 600$^{\circ}$C.

N-type doping is achieved in these films by flowing diluted silane into the chamber during the growth and doping in the range of 2$\times$10$^{16}$ - 2$\times$10$^{19}$ cm$^{-3}$ were studied. To understand the incorporation of Si and other unintentional impurities (C, H) in the films grown at 600$^{\circ}$C, secondary-ion mass spectroscopy (SIMS) measurements were done on a stacked structure with different silane flows (see supplementary information). Figure 2 shows the measured Si atom density (left Y axis) as a function of the silane flow. The Si atom density scales linearly with the silane flow suggesting complete incorporation. Impurities such as H and C are always near the SIMS detection limits at lower silane flows. Thus, high purity $\beta$-Ga$_2$O$_3$ thin films with controllable doping are achieved and are further verified by the excellent transport properties as discussed later. 

\begin{figure}
\centering
  \includegraphics[width=3in,height=5.5cm, keepaspectratio]{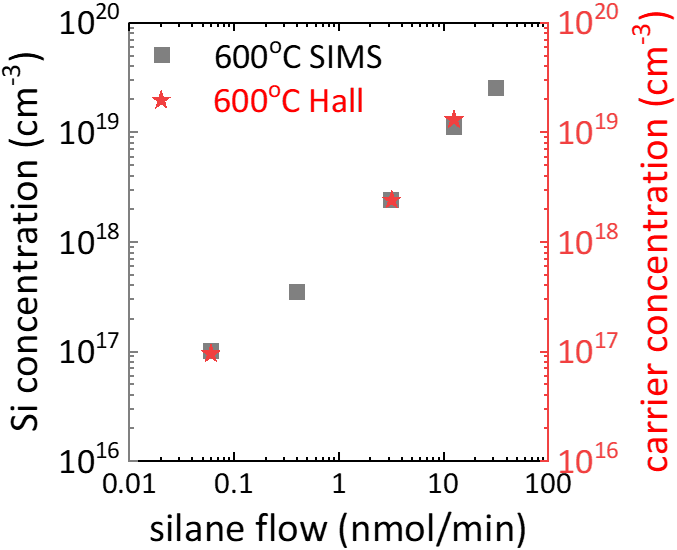}
\label{fig 2}
\caption{Comparison of Si density from SIMS and carrier density extracted from Hall measurements in the  600$^{\circ}$C grown $\beta$-Ga$_2$O$_3$ thin films as a function of silane flow.}
\vspace{-0.15cm}
\end{figure}

\begin{figure}
\centering
\hspace{-0.9cm}
  \includegraphics[width=3in,height=5.5cm, keepaspectratio]{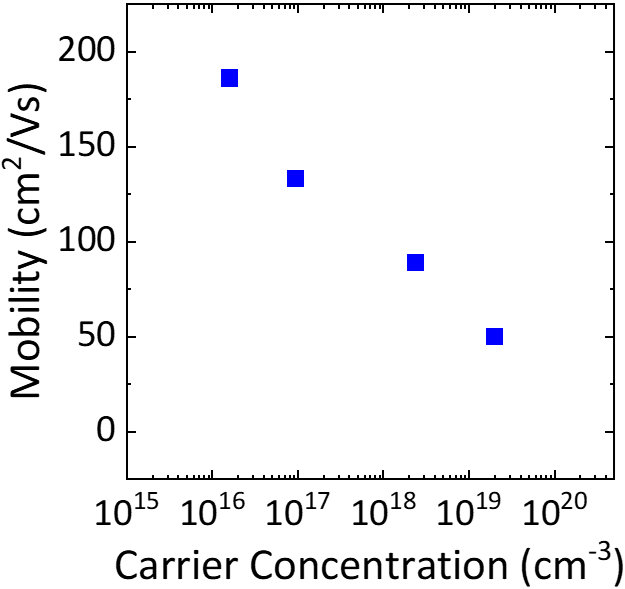}
\label{fig 3}
\caption{Room temperature Hall mobility vs carrier density in the 600$^{\circ}$C grown $\beta$-Ga$_2$O$_3$ thin films. }
\vspace{-0.15cm}
\end{figure}

Hall effect measurements (Ecopia HMS 7000) were performed to study the transport properties of these films. Ohmic contacts were formed by depositing Ti (50 nm)/Au (100 nm) on the four corners of as grown unintentionally doped (UID) and  Si-doped $\beta$-Ga$_2$O$_3$ thin films using DC sputtering. The contacts were further annealed in N$_2$ at 450$^{\circ}$C for 1.5 minutes so that the contacts retained their ohmic nature even at low temperatures. Figure 2 (right Y axis) shows the measured carrier densities as a function of the silane flow for the films grown at 600$^{\circ}$C. The carrier density matches closely with the measured Si density from SIMS which indicates complete activation of the Si atoms. The doping profiles were further extracted by CV profiling (see supplementary) and are found to be very uniform throughout the film thicknesses. Room temperature (RT) Hall mobilities as a function of carrier density is shown in Figure 3. The UID film grown at 600$^{\circ}$C exhibits an impressive room temperature mobility of 186 cm$^2$/Vs (for N$_D^+$ - N$_A^-$ $\sim$ 2$\times$10$^{16}$ cm$^{-3}$)  which is close to the highest reported mobility values of high-temperature MOVPE-grown UID films with similar background doping. This clearly indicates that our LT-grown film is of high crystalline quality (low defects/compensation) equivalent to those grown at high temperatures.

\begin{figure}
\centering
\begin{subfigure}
 \centering
  \includegraphics[width=3in,height=5.7cm, keepaspectratio]{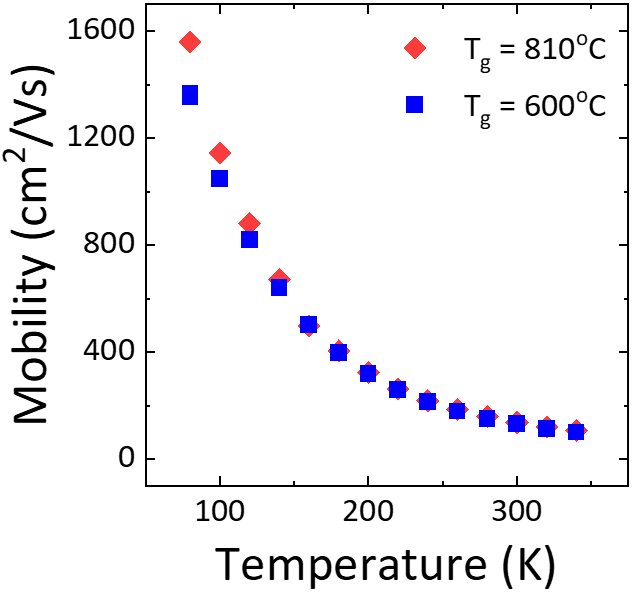}
  \label{fig4a}
\end{subfigure}
\begin{subfigure}
  \centering
  \hspace{2.95cm}
  \includegraphics[width=3in,height=5.5cm,keepaspectratio]{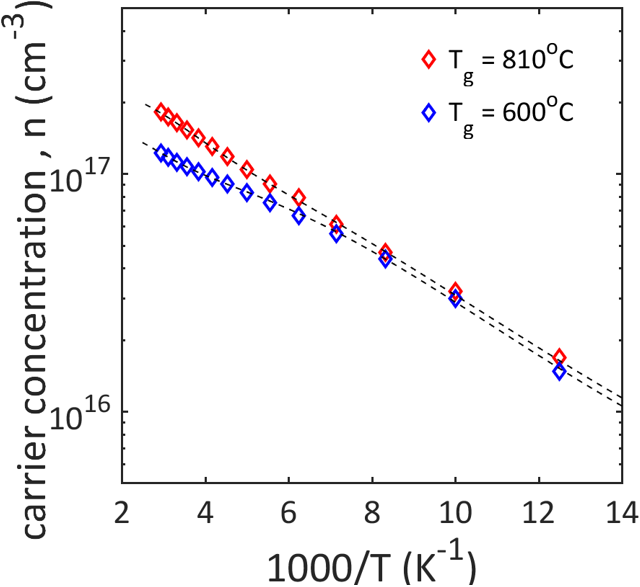}
\end{subfigure}
\label{fig4}
\caption{(a) Temperature dependent Hall mobility of a lightly-doped $\beta$-Ga$_2$O$_3$ thin film grown at 600$^{\circ}$C and compared with a film grown at 810$^{\circ}$C having nominally the same growth conditions. (b) Hall carrier density vs 1000/T of these two films. (Colored diamonds represent the measured values and the black dashed lines represent the fitting to the 2 donor model.) }
\vspace{-0.15cm}
\end{figure}

\begin{figure*}[ht]
\includegraphics[width=7.2in,height=5.7cm, keepaspectratio]{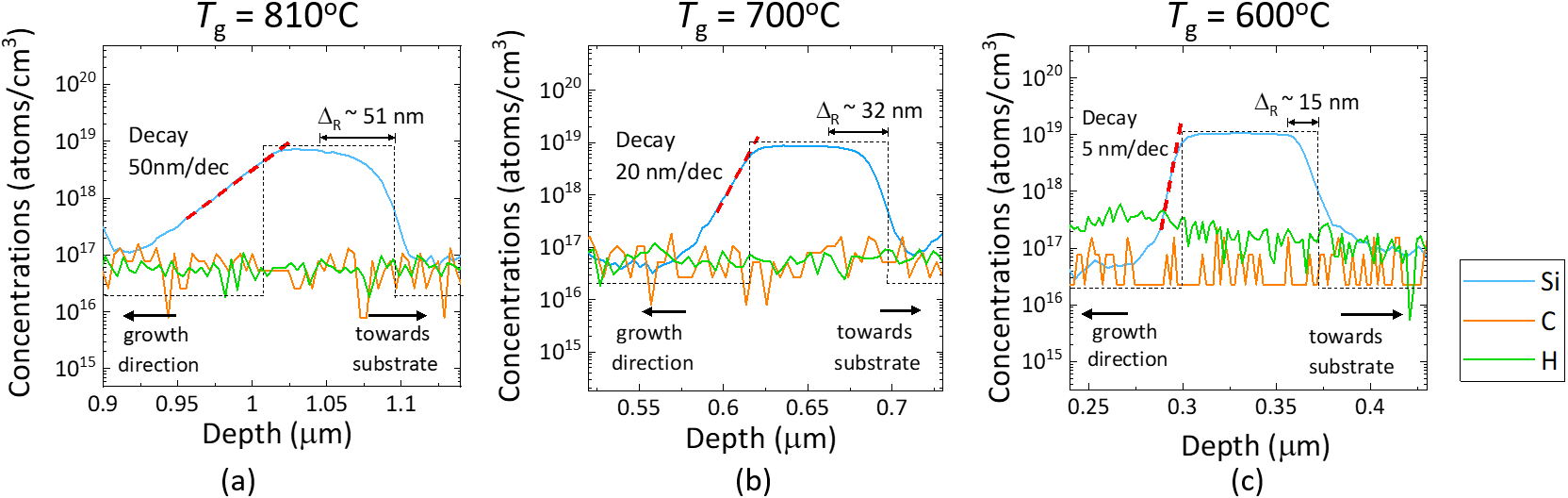}
\caption{Profile of Si, C and H in the films grown at  (a) 810$^{\circ}$C, (b) 700$^{\circ}$C and (c) 600$^{\circ}$C analyzed by SIMS measurements. Black dashed lines indicate the silane switching times and the expected ideal doping profile.}
\vspace{-0.25cm}
\end{figure*}

To further quantify the compensation and the Si donor behavior for films grown at 600$^{\circ}$C, temperature-dependent Hall measurements (80K-340K) were performed and was compared to films grown at 810$^{\circ}$C. Lightly doped films were grown at 600$^{\circ}$C and 810$^{\circ}$C by keeping the silane flows and all other growth conditions nominally the same. Figure 4(a) shows the temperature-dependent mobility of the two films. Both films exhibit high RT mobilities of 140 cm$^2$/Vs (810$^{\circ}$C) and 133 cm$^2$/Vs (600$^{\circ}$C) for carrier densities of 1.75$\times$10$^{17}$ cm$^{-3}$ and 1.2$\times$10$^{17}$ cm$^{-3}$ respectively. The lower mobility for the LT grown film could be due to higher Si contaminants at the epilayer/substrate interface which forms a low mobility parasitic channel \cite{Feng2020}. For low-doped films the contribution from this parasitic channel could be non-trivial \cite{Feng2020}. The effect of this is also evident from the lower LT peak mobility of the LT grown film.

\begin{table}

\captionsetup[table]{
    justification=centering,
    textformat=down,
}
\footnotesize
\caption{\label{tab:table1}Summary of free parameters extracted by fitting the two donor one acceptor charge model to the experimental values. N$_{d1}$ and N$_{d2}$ are the concentrations of the two donors, N$_{a}$ the concentration of compensating acceptors, E$_{d1}$ and E$_{d2}$ the donor energies. An effective mass of m$^*$= 0.28m$_o$ was used to estimate N$_c$ analytically.}
\begin{tabular}{p{0.4cm}p{1cm}p{0.7cm}p{1cm}p{0.7cm}p{1.2cm}p{1cm}}
\hspace{0.03cm} &\hspace{0.03cm}&\hspace{0.03cm}&\hspace{0.03cm}&\hspace{0.03cm}&\hspace{0.03cm}&\hspace{0.03cm}\\
T$_g$ & N$_{d1}$ & E$_{d1}$  & N$_{d2}$  & E$_{d2}$ &  N$_{a}$ & $\mu$ (RT) \\
 ($^{\circ}$C)     & (cm$^{-3}$) & (meV) & (cm$^{-3}$) &  (meV) &   (cm$^{-3}$) & (cm$^{2}$/Vs)\\

\hline
\\
600 & 1.1$\times$10$^{17}$ & 19 & 5.5$\times$10$^{16}$ & 100 & 2.5$\times$10$^{15}$  & 133\\
\\
810 & 1.3$\times$10$^{17}$ & 17 & 7.7$\times$10$^{16}$ & 65 & 2.0$\times$10$^{15}$ & 140\\
\\
\hline

\end{tabular}

\vspace{-0.2cm}
\end{table}

A two donor and one acceptor charge model\cite{neal2018donors, neal2017incomplete} was used to analyze the temperature-dependent carrier densities in the two films as shown in Figure 4(b) and the extracted parameters are summarized in Table I. Both films show a shallow donor state with an activation energy of 17-19 meV that could be attributed to Si. Two different deep donor states are also observed in these films. These donor energy levels have also been observed in films grown by different growth techniques but their origin still remains to be understood. Deep donor states in $\beta$-Ga$_2$O$_3$ could originate from extrinsic traps/defects such as interstitials, antisites and the incorporation of unintentional impurities\cite{neal2017incomplete}. Also, the presence of residual Si impurities at the epilayer/substrate interface complicates the analysis of charge and transport characteristics. To date, no standard processing technique has been developed that could remove the accumulated residual Si impurities which  makes it an  uncontrolled parameter. Nevertheless, the extracted concentrations of compensating acceptors in both the films are very low of the order $\sim$ 2$\times$10$^{15}$ cm$^{-3}$. This shows that $\beta$-Ga$_2$O$_3$ thin films with high crystalline quality (low defects) can now be grown over a wider growth temperature window of $>$200$^{\circ}$C using the MOVPE growth technique. 

Next, we studied the effect of the growth temperature on abruptness of the doping profiles in these Si-doped $\beta$-Ga$_2$O$_3$ thin films. Stacks with intentionally doped layers sandwiched between UID layers (see supplementary information) were grown at different temperatures and only the silane flow was switched without any growth interruption between the layers. Both doping and depth profiles were analyzed by SIMS measurements. Figure 5 shows the doping vs depth profiles of the representative structures with similar silane flows grown at three different temperatures of $T_g$ = 810$^{\circ}$C, 700$^{\circ}$C and 600$^{\circ}$C. It can be seen that the forward Si decay (towards the growth direction) is strongly dependent on the growth temperature and becomes steeper when the growth temperature is lowered. By reducing the growth temperature from 810$^{\circ}$C to 600$^{\circ}$C, the forward Si decay changed from $\sim$ 50nm/dec to $\sim$ 5nm/dec. Note that the decay of the forward Si tail improved monotonically as the growth temperature was lowered. The corresponding forward Si decays for three different growth temperatures are found to be consistent over a wide range of Si densities in the doped layer - 3$\times$10$^{17}$ cm$^{-3}$ to 2$\times$10$^{19}$ cm$^{-3}$ (Figure 6). 

\begin{figure}[h]
\centering
\includegraphics[width=3in,height=5.5cm, keepaspectratio]{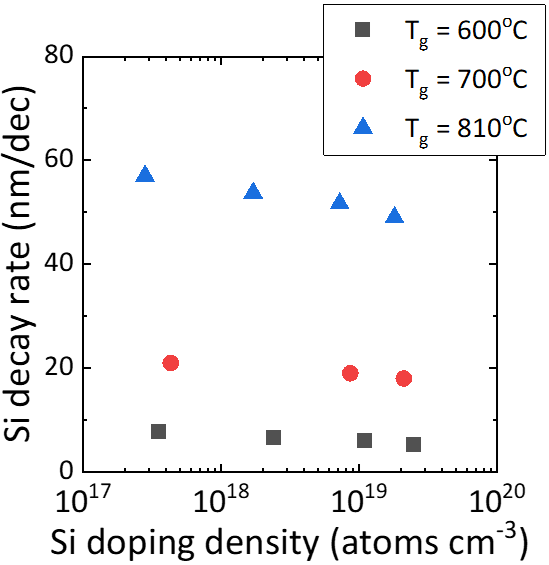}
\label{fig6}
\caption{Forward Si decay rates for various doping densities in films grown at different temperatures.}
\vspace{-0.15cm}
\end{figure}

It was observed that the reverse Si decay (towards the substrate) was nearly independent of the growth temperature as well as the Si density in the doped layer. This suggests that the forward Si tail that extends into the undoped layers could not be explained due to Si diffusion alone \cite{mauze2019investigation, iyer1981sharp}. A possible explanation could be that at higher growth temperatures, a significant amount of Si segregates and accumulates at the growth surface with lower incorporation in the preceding monolayers and continues to "ride" the surface even after the silane flow is swithced off \cite{mauze2019investigation} . This understanding is further strengthened by the observation of a significant increment of spatial delay in reaching the steady-state doping with higher growth temperatures after the silane flow is switched on as shown in Figure 5. The characteristic delay length ($\Delta_R$) is found to decrease as the growth temperature is reduced. This kind of doping profile smearing has been observed in other semiconductor growths as well and has been attributed to variation in dopant segregation, desorption, and incorporation rates with growth temperature \cite{sundgren1989dopant, greene1985modeling, rockett1986si, barnett1985si, kohler2013diffusion, tomita2008reduction,elleuch2019highly}. In our case, the effect of growth temperature on Si segregation at the growth surface could be more pronounced than Si desorption. In general, the incorporation probability during semiconductor growth not only depends on the dopant desorption rates, dopant surface segregation (surface coverages), the incoming dopant flux but also on the growth rate. Some additional strategies such as growth interruption coupled with purge cycles could be useful to achieve even sharper doping profiles. To quantitatively explain the doping profile smearing observed at elevated growth temperatures, modeling and experimental verification of dopant desorption, incorporation, segregation and bulk diffusion for MOVPE $\beta$-Ga$_2$O$_3$ growth are necessary.

In conclusion, we demonstrate the growth of UID and Si-doped $\beta$-Ga$_2$O$_3$ thin films on Fe-doped (010) bulk substrates using the MOVPE technique at a growth temperature of 600$^{\circ}$C which is significantly lower than the conventional MOVPE growth temperatures reported. High purity films with excellent transport properties are achieved evident from the high RT Hall mobility of 186 cm$^2$/Vs measured on UID films as well high mobilities in Si-doped film. Extremely smooth films with atomically flat surfaces are achieved at this growth temperature that was attributed to higher Ga adatom mobility on a Ga$_2$O$_3$ surface. Si density from SIMS and carrier density from Hall measurements matches closely and scales linearly with the silane flow indicating complete incorporation and activation of Si. Compensating acceptors and other impurities such as C and H  are found to be low and comparable to films grown at higher temperatures. Abrupt doping profiles are achieved with forward Si decay as sharp as 5nm/dec. This was achieved due to the suppression of Si "surface riding" by growing at a lower temperature. This demonstration widens the MOVPE growth temperature window for $\beta$-Ga$_2$O$_3$ homoepitaxy by  more than 200$^{\circ}$C which could be beneficial for the development of $\beta$-Ga$_2$O$_3$ based high-frequency power devices.

\section{Acknowledgement}
This work was supported by the Air Force Office of Scientific Research under award number FA9550-18-1-0507 (Program Manager: Dr. Ali Sayir). We also acknowledge the II-VI foundation Block Gift Program for financial support. This work was performed in part at the Utah Nanofab sponsored by the College of Engineering and the Office of the Vice President for Research.

\nocite{*}
\bibliographystyle{IEEEtran}
\bibliography{ltepitaxy.bbl}

\end{document}